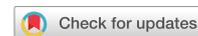

ARTICLE   OPEN

# Donor-acceptor pairs in wide-bandgap semiconductors for quantum technology applications

Anil Bilgin[1 ✉], Ian N. Hammock[1], Jeremy Estes[1], Yu Jin[2], Hannes Bernien[1], Alexander A. High[1,3] and Giulia Galli[1,2,3 ✉]

We propose a quantum science platform utilizing the dipole-dipole coupling between donor-acceptor pairs (DAPs) in wide bandgap semiconductors to realize optically controllable, long-range interactions between defects in the solid state. We carry out calculations based on density functional theory (DFT) to investigate the electronic structure and interactions of DAPs formed by various substitutional point-defects in diamond and silicon carbide (SiC). We determine the most stable charge states and evaluate zero phonon lines using constrained DFT and compare our results with those of simple donor-acceptor pair (DAP) models. We show that polarization differences between ground and excited states lead to unusually large electric dipole moments for several DAPs in diamond and SiC. We predict photoluminescence spectra for selected substitutional atoms and show that while B-N pairs in diamond are challenging to control due to their large electron-phonon coupling, DAPs in SiC, especially Al-N pairs, are suitable candidates to realize long-range optically controllable interactions.

## INTRODUCTION

The search for a quantum architecture that can be easily fabricated, addressed, and manipulated has been at the forefront of academic, industrial, and defense efforts[1]. Realizing such an architecture will enable us to engineer quantum technologies. Diamond and silicon carbide (SiC) have been demonstrated to be suitable platforms to engineer devices for quantum teleportation[2], quantum networks[3], and quantum memories[4,5]. The promise of such solid state systems is driven by the possibility of harnessing spin-spin interactions between color centers and their surrounding nuclear spins, and additionally by establishing entangled photonic links[3,4,6]. However, current implementations of solid-state qubits are limited to the length scale (several angstroms) of spin-spin interactions, or to macroscopic distances of meters to kilometers associated with quantum networks[4,7]. Both of these implementations have several drawbacks. The former relies upon stochastic formation of color centers and nuclear sites, and at present, nanofabrication techniques cannot engineer devices on the angstrom scale[8]. On the other hand, large scale network platforms require entanglement protocols based on probabilistic photon detections that often occur on time-scales much longer than the qubit's coherence time[9,10]. These shortcomings may be overcome with the introduction of a system possessing coherent, long-range (> 10 nm) interactions, and addressable optical transitions. Long-range interactions within the solid-state are particularly interesting as they would enable desirable on-chip entanglement schemes and protocols to improve state-of-the-art methods[11–13].

Many promising avenues have been explored over the years in search of interactions stronger than spin-spin interactions. For instance, in cuprous oxide, manipulation of giant Rydberg excitons with principal quantum numbers of up to $n = 25$ has been demonstrated[14]. The strong, long-range dipole-dipole interactions from these excitons could be used to create solid-state analogs of Rydberg atoms[15]. Unfortunately, it has been difficult to produce high crystal quality artificial cuprous oxide and the observation of such higher excited states is made possible by the comparatively large Rydberg energy of around 100 meV[14]. The specific conditions required for the realization of this class of systems might hamper their generalization to other host materials. Another interesting class of systems are donor-acceptor pairs (DAPs) in semiconductors. The transition between ionized donors and acceptors has been theoretically modeled and experimentally observed in many materials, including silicon[16,17], silicon carbide[18–20], diamond[21,22], and other compound semiconductors[23–26]. Even though donor-acceptor pair (DAP) systems have been extensively studied for decades, only recently they have been considered for applications in quantum information science. For instance, efficient DAP emission in hexagonal boron nitride has recently been proposed as an alternative way of generating single photon sources for quantum technologies[27]. Others have tried assembling DAPs with well-defined donor-acceptor distances and orientations in covalently-linked organic molecules as a way of manipulating and controlling spin states[28,29]. However, compared to molecular systems, DAPs in the solid-state have the advantage of making use of mature nanofabrication and control techniques for implementation on a chip in a scalable manner.

One crucial property of DAPs that could allow the realization of emergent quantum phenomena in solids is their static electric dipole moments that can be optically controlled, similar to Rydberg atoms in vacuum. This dipole is due to the donor and the acceptor remaining charged after the electron and hole pair recombine to produce a photon. A pictorial description of the electronic ground and excited states of DAPs is given in Fig. 1c. The strength of the dipole depends on the distance between the donor and the acceptor, and in the future, it might be engineered by placing the impurities at specific distances in the host lattice, as illustrated in Fig. 1a, which requires substantial developments in nanofabrication techniques.

Here, we propose a quantum science platform utilizing the static electric dipole-dipole coupling between DAPs in wide bandgap semiconductors to realize optically controllable, long-range interactions in the solid-state. The host crystal could then be interfaced with devices controlled at the nanoscale, using nanofabrication techniques. In our proposed platform, pair-pair interactions are mediated by the optically switchable dipole-dipole coupling similar to Rydberg interactions in atomic systems[30,31]. This coherent interaction is governed by Coulomb's law, and depends on the pair-pair distance and orientation. Hence it can extend over a significantly longer length scale (>10 nm) than spin-spin interactions. With emerging developments in growth and synthesis techniques[8], it might be possible to create arrays of DAPs with well-defined pair distances and orientations, similar to molecular systems. Additionally, DAPs in solid-state systems have the advantage of easily-fabricated control infrastructures such as cavities and waveguides. Moreover, DAP lifetimes of up to tens of microseconds[17], or even up to milliseconds[32] have already been observed. These lifetimes are much longer than the hundreds of nanoseconds observed for charge recombination in molecular DAP systems[33], paving the way to realizing long-lived states that can be coherently controlled and utilized for quantum applications. The broad variety of DAPs available in different host materials already used for quantum information science highlights the general applicability of the proposed platform.

[1]Pritzker School of Molecular Engineering, University of Chicago, Chicago, IL 60637, USA. [2]Department of Chemistry, University of Chicago, Chicago, IL 60637, USA. [3]Center for Molecular Engineering and Materials Science Division, Argonne National Laboratory, Lemont, IL 60439, USA. ✉email: bilgin@uchicago.edu; gagalli@uchicago.edu

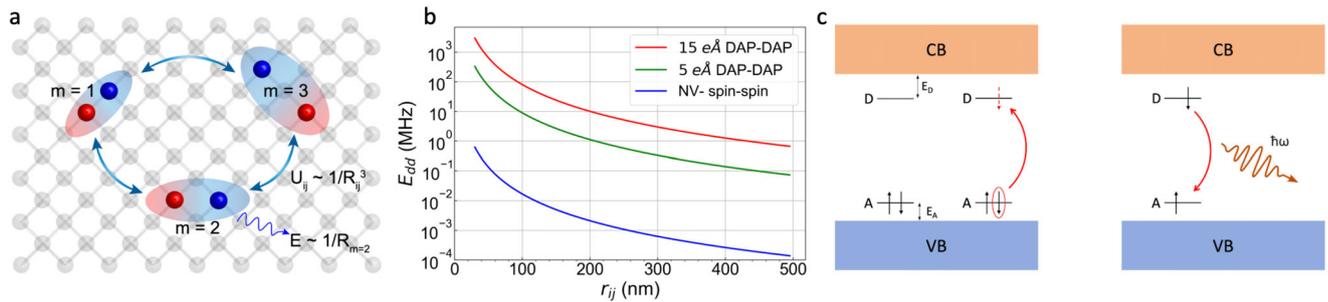

**Fig. 1 Donor-acceptor pair interactions and electronic structure. a** An illustration of donor-acceptor pairs (DAPs) where the donor and acceptor are first (m = 1), second (m = 2) and third (m = 3) nearest neighbors. The dipole-dipole interaction between DAPs scale as the inverse cube of the distance between the pairs ($R_{ij}$). The photon energy emitted from a given shell (m) is proportional to the inverse pair distance ($R_m$) between the donor and acceptor of that shell. **b** The dipole-dipole interaction strength of two DAPs assumed to have dipole moments of 15 eÅ (red, expected dipole moment from shell m45) and 5 eÅ (green, expected dipole moment from shell m6) in 3C-SiC, compared to spin-spin interactions of NV- centers in diamond (blue). **c** Electronic structure of a donor (D) acceptor (A) pair in the ground (left) and excited (right) states. $E_D$ and $E_A$ are the donor and acceptor binding energies and VB and CB denote valence band and conduction band of the host material, respectively. A photon is emitted as a result of the recombination of a hole bound to the acceptor and an electron bound to the donor.

We present first-principles calculations of the electronic structure of DAPs to understand some of the key properties that are crucial to realize the proposed platform experimentally. We focus our efforts specifically on silicon carbide and diamond, though many other semiconductors may be viable, and the methods presented here are transferable to other systems. First, in order to evaluate whether DAPs could facilitate long-range quantum interactions in the solid-state, we investigate the electronic structure of DAPs, specifically the corresponding electronic states in the bandgap of their host materials. We compute the charge transition levels (CTL) for several donors and acceptors and calculate their binding energies. Then, we determine the zero-phonon lines (ZPL) of specific DAPs and predict their photoluminescence (PL) spectra to assess whether individual ZPL from increasingly distant pairs can be experimentally resolved. We compute the magnitude of electric dipole moments that enable DAPs to have optically controllable long-range interactions to show that strong interactions at longer than 10 nm length scales can indeed be realized. Finally, we provide estimates of the order of magnitude of the radiative lifetimes of some of the DAPs studied here. Our findings indicate that DAPs composed of shallow donors and acceptors exhibit weaker electron-phonon coupling and hence are easier to resolve in photoluminescence spectra. We propose that shallow DAPs may be used as building blocks for realizing an optically addressable, strongly interacting quantum system in the solid-state.

## RESULTS
### Charge transition levels
We analyzed several substitutional defects: boron ($B_C$), nitrogen ($N_C$) and phosphorus ($P_C$) defects in diamond as well as boron ($B_C$), nitrogen ($N_C$) replacing a carbon atom and aluminum ($Al_{Si}$) replacing a silicon atom in SiC. We first computed the charge transition levels ($E^f[X^q]$) of all defects ($X$) as a function of the charge state $q$, with $q = 0$ and $q = 1(q = -1)$ for donors (acceptors)[34]:

$$E^f[X^q] = E_{tot}[X^q] - E_{tot}[bulk] - \sum_i n_i \mu_i + qE_F + E_{corr}, \quad (1)$$

where $E_{tot}[X^q]$ and $E_{tot}[bulk]$ are the total energies of the supercells containing the defect $X$ and of the pristine bulk, respectively. The integer $n_i$ indicates the number of atoms of type $i$ (host or defect) that are added or removed from the bulk supercell to create the defect, $\mu_i$ are the corresponding chemical potentials and $E_F$ is the Fermi level. The term $E_{corr}$ is a correction that accounts for the fictitious electrostatic interaction between supercells[35] (see Supplemental Information Section A and F for details of the calculations).

Our results are reported in Table 1. Calculations using the HSE[36] functional yield results in good agreement with those of previous studies, that are within 0.1 eV of experiment. Note the difference between results obtained with the PBE[37] and HSE functionals in the case of the $N_C$ donor in diamond, mostly due to the different geometrical distortions predicted by the two functionals. Interestingly the PBE and HSE functionals perform identically for the $N_C$ shallow donor in 3C-SiC, slightly overestimating its binding energy. The differences between the acceptor levels for $B_C$ and $Al_{Si}$ computed at the PBE and HSE level of theory are again due to differences in geometries (see Supplementary Section A). In order to estimate errors due to finite size effects we carried out calculations with up to 1000 atoms with the PBE functional and we extrapolated our results to the dilute limit (see Supplementary Section F). The error on computed CTL values is shown in the Supplementary Information section and is on the order of 0.05 eV.

### Electric dipole moments
In order to understand whether the DAPs discussed above can be manipulated, we first estimated the magnitude of their electric dipole moments using the simple expression $eR_m$, as shown in

Fig. 2. However, this is an approximation, as it does not explicitly consider the exact electronic charge transferred between the donor and the acceptor. To better evaluate such a transfer, we can compute the electric dipole moments as the difference in polarization ($\Delta P$), on the same branch of the polarization lattice, between the initial (excited) and final (ground) states[38], following refs. [39,40]:

$$\Delta P = \frac{1}{\Omega}\left[\sum_i Z_i R_i - \sum_n^{occ} e\bar{\mathbf{r}}_n\right], \quad (2)$$

where $\Omega$ is the volume of the supercell, $Z_i$ and $R_i$ are the charges and positions of individual nuclei; $e$ is the electron charge, $\bar{\mathbf{r}}_n$ is the center of a maximally localized Wannier function (MLWF)[41,42] and the sum over $n$ runs over all MLWF constructed from the occupied eigenstates of the Kohn-Sham Hamiltonian. We note however, that the computed dipole moment depends on the relative orientation of the distorted bonds at the donor and acceptor sites. For example, if the distorted bonds are facing each other in the crystal, the resulting dipole moment is expected to be less than $eR_m$; if facing away from each other, the dipole moment will be larger than $eR_m$. In experiments, a multitude of orientations of D and A pairs would be measured in a given sample and an average value of the dipole moment would be obtained. We conducted a series of calculations to compute the electric dipole moment of selected DAPs in diamond and SiC as a function of the relative orientation of the D and A, using the MLWF method. We obtained dipole moment values both larger and smaller than $eR_m$ and, in cases where we could afford computing multiple orientations and hence a robust average value, we obtained results close to $eR_m$ (See Fig. 2 for the m4 and m5 shells in diamond, and m4 shell in 3C-SiC).

Our results show that the DAPs electric dipole moments can be rather large (>25 Debye). Comparatively, the nitrogen vacancy center (NV-) has an electric dipole moment of ~0.8 Debye[43]. Interestingly, we find that an interaction strength of ~100 MHz can be felt from up to 100 nm away from the DAP, depending on the strength and orientation of the dipoles involved (see Supplementary Section E). Hence DAPs' large dipole moments make them suitable candidates for providing long-range interactions in the solid state. However, their large dipole moments also make them highly sensitive to electromagnetic noise that can broaden optical transitions. Such noise can be mitigated by integrating DAPs in PIN heterostructures, as experimentally demonstrated, for example, for divacancies in SiC[44]. Our proposed DAP platform is fully compatible with such structures - for instance, they can be included in functional material heterostructures such as PIN structures during growth or through implantation - and given that the achievable dipolar interaction strength for DAPs far exceeds MHz level noise, we expect that robust, coherent control for this system should be achievable.

We now turn to addressing whether the zero-phonon lines (ZPL) of the DAPs may be detectable experimentally and whether ZPLs coming from DAPs located at various distances may be separated from each other.

### Zero phonon lines

We start by analyzing ZPLs as a function of the distance $R_m$ and we show that a simple model can accurately reproduce the results obtained with constrained DFT. We then discuss the sensitivity of ZPLs to electric fields. Using constrained DFT[45,46], the ZPLs were evaluated as the difference between the total energy of the excited state and of the ground state, in their respective optimized geometries, and they were computed as a function of the distance $R_m$ between the donor and the acceptor. We also computed the value of the ZPLs using a DAP model, where the energy $\hbar\omega_m$ as a function of $R_m$ is expressed as[16,23,25]:

$$\hbar\omega_m = E_g - (E_A + E_D) + \frac{k_e e^2}{\varepsilon_r r_b}\frac{r_b}{R_m} + J(R_m), \quad (3)$$

where $E_g$ is the bandgap energy, $E_D$ and $E_A$ are the donor and acceptor binding energies calculated from Eq. (1)[34], $e$ is the

**Table 1.** Charge transition levels.

| Host | CTL | PBE (HSE) | Previous work |
|---|---|---|---|
| Diamond[a] | $B_C$ (0/−) | $E_V + 0.35(0.35)$ | $E_V + [0.37–0.38]$[c] |
| | $N_C$ (+/0) | $E_C − 1.47(1.80)$ | $E_C − [1.7–1.9]$[d] |
| | $P_C$ (+/0) | $E_C − 0.45(0.47)$ | $E_C − [0.5–0.6]$[e] |
| 3C-SiC[b] | $B_C$ (0/−) | $E_V + 0.37(0.57)$ | $E_V + [0.6–0.73]$[f] |
| | $N_C$ (+/0) | $E_C − 0.16(0.16)$ | $E_C − [0.05–0.06]$[g] |
| | $Al_{Si}$ (0/−) | $E_V + 0.17(0.19)$ | $E_V + [0.24–0.26]$[h] |

Computed charge state transition levels (CTL), referenced in eV to the valence ($E_V$) and conduction ($E_C$) band edges, for boron ($B_C$), nitrogen ($N_C$) and phosphorous ($P_C$) in diamond, and boron ($B_C$) and nitrogen ($N_C$) replacing carbon and aluminum replacing silicon ($Al_{Si}$) in 3C-SiC. We report results obtained with two density functionals, PBE[37] and HSE[36] as well as results of previous work.
[a]$E_g = 4.17$ (5.37) eV for the PBE (HSE) functional
[b]$E_g = 1.40$ (2.25) eV for the PBE (HSE) functional
[c]Ref. [71,72]
[d]Ref. [72–77]
[e]Ref. [78,79]
[f]Ref. [18,80]
[g]Ref. [18,81,82]
[h]Ref. [80–82]

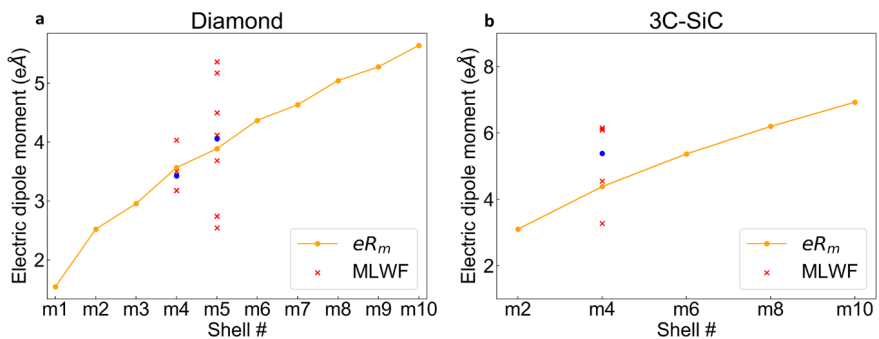

**Fig. 2 Electric dipole moments of donor-acceptor pairs.** Electric dipole moments for B-N pairs in diamond (**a**) and 3C-SiC (**b**) calculated using the approximate formula $eR_m$, where $R_m$ is the donor-acceptor distance (orange dots) and Maximally localized Wannier Functions (MLWF, blue dots). The MLWF values have been obtained by averaging over the different configurations (red crosses) that the distorted bonds may attain in the crystal. The details on the relative orientations of each data point are given in the Supplementary Information section.

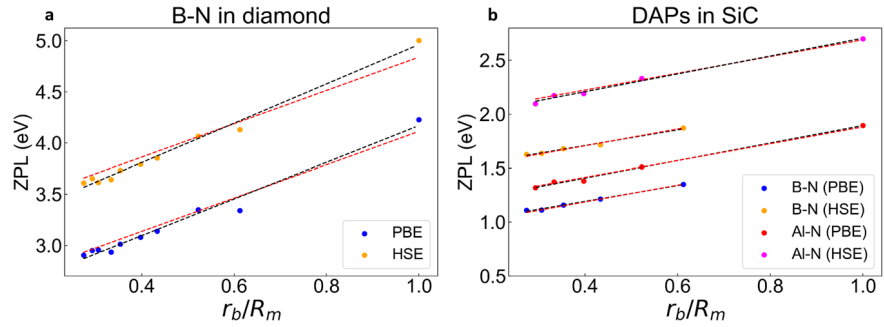

**Fig. 3 Zero-phonon lines of donor-acceptor pairs.** Calculated zero-phonon line (ZPL) energies using constrained DFT are compared with those obtained with the DAP model of Eq. (3) for both diamond (**a**) and 3C-SiC (**b**). $r_b$ is the equilibrium bond length between nearest neighbor atoms (1.55 Å for diamond, 1.90 Å for SiC) and $R_m$ is the distance between the donor and acceptor for a given shell (m). Red dashed lines represent Eq. (3) and the black dashed lines are the results obtained from constrained DFT.

electronic charge, $\varepsilon_r$ is the long-wavelength dielectric constant, and $r_b$ is the equilibrium bond length between nearest neighbor atoms of the host material ($r_b = 1.90$ (1.55) Å for SiC (diamond)). The term $J(R_m)$ is given by[24,47]:

$$J(R_m) = \frac{e^2}{4\pi\varepsilon_0\varepsilon_r} \iint_V \psi_D^*(\mathbf{r})\psi_A^*(\mathbf{r}')\left[\frac{1}{|\mathbf{r}-\mathbf{R_m}|} + \frac{1}{|\mathbf{r}'+\mathbf{R_m}|}\right.$$
$$\left. - \frac{1}{|\mathbf{r}-\mathbf{R_m}-\mathbf{r}'|} - \frac{1}{|\mathbf{R_m}|}\right]\psi_D(\mathbf{r})\psi_A(\mathbf{r}')d^3\mathbf{r}d^3\mathbf{r}'. \quad (4)$$

Here, $\psi_D$ and $\psi_A$ are the wavefunctions for the electron and hole bound to the donor and the acceptor, respectively (see Supplementary Section C).

Our results obtained with constrained DFT and the DAP model are reported in Fig. 3 and summarized in the Supplementary Information section. The behavior of ZPL as a function of increasing shell number is well reproduced by the DAP model. In particular, the slopes obtained from computed ZPL values for both Al-N and B-N pairs in 3C-SiC are close to the ones predicted by the model. The PBE and HSE functionals yield similar results for B-N and Al-N pairs in 3C-SiC. For B-N DAPs in diamond, the prediction for the slope by the model differs considerably compared to other DAPs in 3C-SiC. This difference stems from the approximation used for the wavefunctions in the DAP model (perturbed Hydrogen-like wavefunctions[24,47]), which has different ranges of validity depending on the host and pair. In particular, such an approximation performs less well for defects that lie deeper in the bandgap, such as the $N_C$ donor in diamond.

From the y-intercept in Fig. 3, we can infer the sum of the binding energies of the donor and acceptor. In all cases, the predicted sums for the binding energies are within 0.1 eV of the calculated combinations of sums of binding energies from Table 1. Although we obtained all of our results from first principles, our comparisons with the DAP model show that this model is a useful tool to predict ZPL energies as a function of distance and can provide (semi-)quantitative results. Our calculated ZPL energies for the first few shells of Al-N pairs in 3C-SiC using the HSE functional are in the range 2.10–2.25 eV, in agreement with experimental values in the range 2.25–2.30 eV[19,20]. Similarly, our HSE predictions for the first few shells of B-N pairs in 3C-SiC fall within 1.6–1.8 eV, remarkably close to previous experimental observations in the range 1.75–1.95 eV[18].

Next we investigate how individual DAPs respond to applied electric fields. As an example, we report in Fig. 4 the computed ZPL of the m5 shell in diamond, which shows an extraordinary sensitivity to applied electric fields, with a tunability of the ZPL possible over a few THz. This extreme tunability of DAPs can be leveraged as an electronically tunable source of single photons as well. The Stark shift for the ZPL of DAPs responsible for this

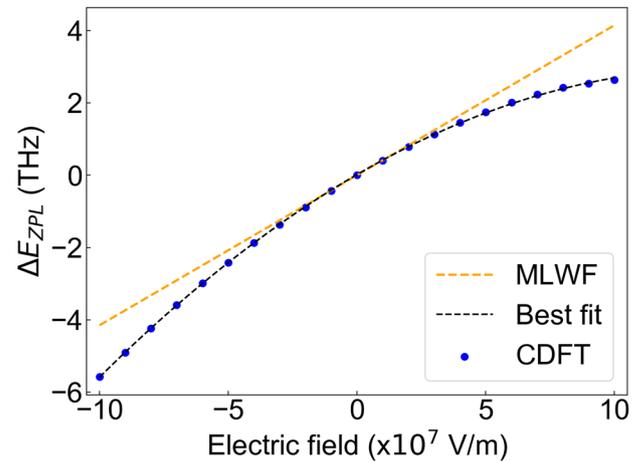

**Fig. 4 Stark shift of donor-acceptor pairs.** Computed zero phonon line shift ($\Delta E_{ZPL}$) for the B-N m5 shell in diamond under homogeneous applied electric field in the (100) direction. The slope at zero field gives the polarization (which can be converted to dipole moment). The orange dashed line indicates the first-order energy shift to ZPL with the slope obtained from maximally localized Wannier functions (MLWF) calculations. The dark blue circles are results obtained from constrained DFT calculations with the PBE functional and the dashed line is a quadratic fit.

tunability can be expressed as:

$$\Delta E_{ZPL} = -\Delta\boldsymbol{\mu} \cdot \mathbf{E} - \frac{1}{2}\mathbf{E} \cdot \Delta\boldsymbol{\alpha} \cdot \mathbf{E} \quad (5)$$

where $\Delta\boldsymbol{\mu}$ is the electric dipole moment, $\Delta\boldsymbol{\alpha}$ is the polarizability tensor and $\mathbf{E}$ is the applied electric field. Incidentally, the electric dipole moment of the DAP can be extracted from the instantaneous slope where the applied field is zero, and we find values that compare very well with those obtained above by averaging our results computed with MLWFs (see Supplementary Section B).

### Electron-phonon coupling and photoluminescence spectra

So far we have demonstrated that DAPs exhibit large electric dipole moments capable of providing optically controllable long-range interactions, and we have shown the sensitivity of ZPLs to electric fields. Next, we investigated whether, in a given host, DAPs at different distances may be optically resolved in experiments. To this end, we compute the photoluminescence (PL) spectra using the effective one-dimensional (1D) configurational coordinate (CC) approximation[48–52]. Within this model, the

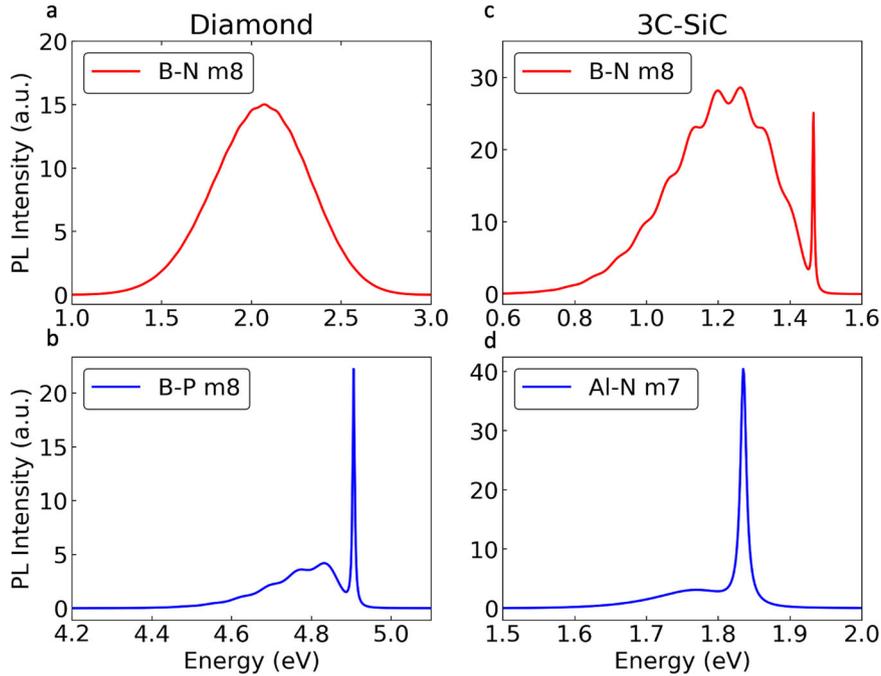

**Fig. 5 Photoluminescence spectra of donor-acceptor pairs.** Photoluminescence (PL) spectra obtained from one-dimensional configurational-coordinate diagram using the HSE functional for (**a**) B-N (m8) and (**b**) B-P (m8) pairs in diamond, and (**c**) B-N (m8) and (**d**) Al-N (m7) pairs in 3C-SiC. PL lineshapes were calculated at 5K using a Lorentzian broadening parameter $\gamma = 3$ meV for the zero-phonon line (ZPL) and a Gaussian broadening parameter $\sigma = 30$ meV for approximating the broadening of the spectral function $S(\hbar\omega)$ for phonon modes responsible for the phonon side-band (See ref. [83]).

normalized luminescence lineshape $L$ is given by[53]

$$L(\hbar\omega, T) = \sum_{n,m} w_m(T) |\langle \chi_{em} | \chi_{gn} \rangle|^2 \times \delta(E_{ZPL} + \hbar\omega_{em} - \hbar\omega_{gn} - \hbar\omega). \tag{6}$$

The sum is carried over all the vibrational levels with energies $\hbar\omega_{em}$ and $\hbar\omega_{gn}$ for excited and ground states; $w_m(T)$ is the thermal occupation factor and $\chi$ are the ionic wavefunctions. This formulation of the Franck–Condon approximation assumes that the transition dipole moment is independent of the ionic coordinates[53]. Within this framework each normal mode $k$ contributes to the geometrical distortion of the ions around the defect, with weight $p_k = (\Delta Q_k / \Delta Q)^2$, where

$$\Delta Q_k = \sum_{ai} m_a^{1/2} \Delta R_{ai} q_{k;ai}; \quad (\Delta Q)^2 = \sum_k \Delta Q_k^2, \tag{7}$$

where $a$ indexes run over all atoms, $i = \{x, y, z\}$, $\Delta R_{ai} = R_{e,ai} - R_{g,ai}$ is the vector connecting the ground and excited state geometries; $q_{k,ai}$ is the unit vector in the direction of the normal mode $k$. We define effective frequencies associated to the ground and excited states as:

$$\Omega^2_{\{e,g\}} = \langle \omega^2_{\{e,g\}} \rangle = \sum_k p_{\{e,g\};k} \omega^2_{\{e,g\};k} \tag{8}$$

$$S_{\{e,g\}} = \frac{\Omega_{\{e,g\}} \Delta Q^2}{2\hbar}, \tag{9}$$

where $\omega_{\{e,g\};k}$ is the frequency of mode $q_{\{e,g\};k}$ and $S_{\{e,g\}}$ are the effective Huang–Rhys (HR) factors for the ground and excited states[54]. The magnitude of the HR factors gives an indication of the strength of the electron-phonon coupling in the system: the smaller the electron-phonon coupling, the easier it should be to distinguish the signal of DAPs located at different distances (i.e. for different shells). We show the calculated parameters from the one-dimensional model and report a summary of the results in the Supplementary Information section. B-N pairs in diamond exhibit a very large HR coupling due to the deep nature of the $N_C$ donor. The large lattice distortion that $N_C$ undergoes upon changing its charge state from positive to neutral causes this transition to have a large $\Delta Q$, and a large HR coupling. In comparison, B-P pairs have a lower HR coupling, because of the smaller lattice distortions that $P_C$ experiences upon a charge state transition. As can be seen from Table 1 and the Supplementary Information section, $Al_{Si}$ is a shallower defect compared to $B_C$ and it undergoes a smaller geometric distortion when changing its charge state; the same is true of the comparison between $P_C$ and $N_C$ in diamond.

There are some pronounced differences between the electron-phonon coupling predicted by the PBE and HSE functionals. The total mass-weighted distortions between the ground and excited states ($\Delta Q$) are always larger with the HSE functional. Indeed, PBE predicts most defects to be shallower than HSE and experiment, and this is reflected in the differences between geometric distortions. Hence we conclude that the HSE functional captures the DAP physics more accurately than PBE for two reasons: the position of defect levels within the band gap and the associated lattice distortions upon change of the defect's charge state.

We present in Fig. 5 the photoluminescence spectra for the eighth shell (m8) of B-N and B-P pairs in diamond as well as eighth shell (m8) for B-N pairs and the seventh shell (m7) for Al-N pairs in 3C-SiC using the HSE functional. We find that the B-N pairs in diamond exhibit the largest amount of HR coupling (~20), resulting in a broadband lineshape for the PL spectra, rendering their ZPL not resolvable. In all three other cases, the ZPL appears to be resolvable. However, B-N pairs in SiC are still a challenging case, as the phonon sidebands from multiple shells overlap and create a broad background. Instead, multiple shells of Al-N pairs in 3C-SiC or B-P pairs in diamond should be distinguishable, given their small phonon sidebands, even if the energy differences between consecutive shells are as small as 5–10 meV. With the exception of B-N pairs in diamond, our predictions agree with previous experimental observations of sharp line spectra in 3C-SiC both for Al-N[19,20] and for B-N pairs[18].

Based on previous work[21,55], we measured the photoluminescence spectra of eleven diamond samples containing nitrogen and boron, with different grades and preparation. All samples were measured at 4K, with a 407 nm excitation. We selected two samples to fabricate electrostatic gates on the surface. Sample details can be found in the Supplementary Information section. From previous observations[21,55] we anticipated to see many resolvable B-N DAP peaks between 450 nm and 600 nm. We did not observe this in any sample. Moreover, the electrostatic gate measurements revealed that the defects we did observe, did not possess the large electric dipole moments characteristic of DAPs. In this dataset we realized electric fields on the order of a kV cm$^{-1}$ through our implanted layer and extracted dipole moments less than 0.5 eÅ. These experiments are consistent with our DFT findings: B-N pairs in diamond have a weak and unresolvable ZPL emission due to their strong phonon coupling.

Instead, our results show that DAPs composed of shallow donors and acceptors should be ideal candidates for resonant and off-resonant manipulation, since there is a smaller electron-phonon coupling and shells that are energetically close together can still be experimentally resolved and isolated from each other. Being able to isolate and manipulate DAPs belonging to larger shell numbers comes with the desired availability of increased electric dipole moments, leading to stronger and consequently longer-range interactions in the solid.

### Radiative lifetimes

Finally, we investigated whether the radiative lifetime of DAPs are sufficiently long to allow for sufficient manipulation of their states. Long lifetimes are generally desirable to allow for various pulse sequences to act on the DAP states. We computed radiative lifetimes for DAPs using the expression[56–58]:

$$\tau = \frac{3\epsilon_0 hc^3}{2n_r \omega^3 |\mu|^2}; \quad \mu = \frac{e\langle \psi_f | [\hat{H}_{KS}, \mathbf{r}] | \psi_i \rangle}{\epsilon_f - \epsilon_i} \quad (10)$$

where $\mu$ is the optical transition dipole moment, computed following Eq. (10) in ref. [59], and evaluated in the excited state, where an electron-hole pair is bound to the donor and the acceptor, respectively[58]. $\epsilon_i$ and $\epsilon_f$ are the eigenvalues of the initial and final states $\psi_i$ and $\psi_f$, $\hat{H}_{KS}$ is the Kohn-Sham Hamiltonian, $\mathbf{r}$ is the position operator, $n_r$ is the refractive index of the host material and $\omega$ is the frequency of the optical transition.

In the current literature there is much disagreement about radiative recombination lifetimes of DAPs, depending on the host and defect atoms, ranging from 100s of ps[55] to 10s of μs[16,17] and even few ms[32]. In addition, temperature and dopant concentration levels play an important role[60] in determining lifetimes. Thomas, Hopfield and Augustyniak[60] observe luminescence in GaP after flash excitation, and their model gives μs range results. However, they also observe less than 10$^{-8}$ second lifetimes for certain closely separated pairs[60]. Ziemelis and Parsons[17] attribute their observation of 70–100 μs lifetimes in Si to pairs that are separated by 7.7 to 20 Å. On the other hand, Schneider and Dischler observe 100s of ps of radiative decay times from DAPs in diamond[55].

Our computed radiative recombination lifetimes of DAPs in diamond and 3C-SiC are estimated to be in the ns range for the first ten shells in diamond; these results are closer to the lifetimes obtained by Schneider and Dischler in their experiments and appear to agree with the observation of less than 10$^{-8}$ second lifetimes for closely-separated pairs[55]. Our predictions for lifetimes in 3C-SiC are longer, in part because of the energy difference involved in the transition between charged states. For example, the B-N transitions in diamond are much more energetic than B-N or Al-N transitions in SiC, and the approximately one order of magnitude difference in lifetimes is explained through this energy difference. Another order of magnitude difference comes from the optical transition dipole moment. The squared norm of the optical dipole moment is roughly one order of magnitude smaller in 3C-SiC compared to diamond. Together, the differences in energy level positions and optical dipole moment give rise to sizeable differences between the expected lifetimes of B-N pairs in diamond and 3C-SiC. Our results for calculated radiative lifetimes for B-N pairs in 3C-SiC are in the μs range, which are closer to the predictions of ref. [60]. The longer lifetimes found here for pairs in 3C-SiC are an additional reason why these systems would be more desirable for realizing our proposed platform for optically controlled long-range interactions. Radiative lifetimes that are in the μs range can allow for sufficient manipulation of their states with lasers, whereas ns range lifetimes for B-N pairs in diamond would not suffice. However, additional investigations are required to understand the impact of non-radiative processes.

## DISCUSSION

In this work we proposed that DAPs in wide bandgap semiconductors can be engineered as building blocks for coherent long-range interactions in solids. In this platform, the optically switchable dipole moments of DAPs act as the main mediator of the interactions. We presented a detailed analysis of various DAPs in diamond and 3C-SiC using first-principles calculations, aimed at predicting a few crucial properties of these pairs. We selected a small number of examples, with some deep and shallow defects to illustrate their differences. However, there are many additional defect candidates, and our list of examples is not an exhaustive one. We also restricted our discussion to two host materials, namely diamond and SiC, due to their widespread use in developing quantum technologies. However, the ideas presented here are broadly applicable to other semiconductors, and the fact that DAPs were previously observed in many different semiconductors[16,17,23,26,27] warrants further investigation of those host materials as viable host candidates. Silicon, especially, might be a useful host due to its emission in the telecommunication wavelength range and already-established mature nanofabrication methods for this material. We showed that DAPs exhibit large, optically switchable dipole moments that are capable of providing the basis for long-range interactions (>10 nm) in the crystal. We also showed that the ZPL of DAPs show a broad range of tunability. Our estimates for the radiative recombination lifetime in both host materials show that DAPs can have long radiative lifetimes on the order of μs to allow for sufficient manipulation of these states. Additionally, by using a one-dimensional model, we provided an assessment of the amount of electron-phonon coupling specific DAPs would have. We showed that shallower DAPs exhibit smaller electron-phonon coupling due to a modest rearrangement of their geometries when changing their charge state. Shallow DAPs exhibit line spectra that are not overlapping, allowing for experimental resolution of pairs whose ZPL are separated by just a few meV. The combination of several desirable properties of shallow DAPs in SiC pave the way for coherent manipulations of individual shells and optical control of their long-range dipole-dipole coupling in crystals. Further, our calculations showed that the wavelengths for transitions in DAPs are longer than those for Rydberg atoms, making these transitions more convenient than in Rydberg atoms. In addition, one might envision to fabricate nanophotonic devices such as cavities to efficiently couple light to the DAPs in the future. Regarding the readout of the DAP state, there are potentially multiple strategies. For instance, from the DAP excited state, the defect could be efficiently ionized by further excitations to the conduction band. A subsequent fluorescent image would then reveal what the state of the DAP was. A second approach would use the interaction between two DAPs where the shift on one of them can be probed optically and would reveal the state of the other. Finally, we note that a combination of spin-memory and dipolar interactions might be envisioned, as a mechanism for quantum information storage and processing. Overall the calculations, analysis and predictions reported here suggest that DAPs are a promising platform to engineer optically addressable long range interactions between point-defects in solids for the realization of quantum technologies.

## METHODS
### Details of DFT calculations

We determined the electronic structure of donor acceptor pairs in diamond and 3C-SiC using spin-polarized density functional theory (DFT), and the planewave pseudopotential method as implemented in the Quantum Espresso package[61–63]. The calculations of optical transition dipole moments were carried out using the WEST code[64,65]. We used SG15 ONCV norm-conserving pseudopotentials[66,67] and an energy cutoff of 90 Ry. Calculations of the ZPL's Stark shift under applied electric field were carried out using the PBE semi-local functional[37]. The calculation of electric dipole moments were also based on the PBE functional, and later post-processing involved using MLWFs with the Wannier90 package of Quantum Espresso. All other calculations involved made use of both PBE and the screened hybrid functional HSE06 (HSE)[36], and for those calculations we report results for both functionals. The screening and mixing parameters used for HSE were 0.2 Å$^{-1}$ and 25% respectively. We used a supercell with 512 atoms for all calculations. The lattice parameters ($a_0$) were converged using the PBE (HSE) functional; they are $a_0 = 3.568(3.543)$ Å for diamond and $a_0 = 4.381(4.362)$ Å for 3C-SiC, respectively, compared with the experimental values of 3.567 Å for diamond[68] and 4.360 Å for 3C-SiC[69]. All geometries were relaxed with a force threshold of 1 meV Å$^{-1}$. The Brillouin zone of the supercell was sampled at the Γ point only. Further details of geometry optimizations, zero-phonon lines and electron-phonon calculations are given in the Supplementary Information section.

## DATA AVAILABILITY

Data that support the findings of this study will be made available through the Qresp[70] curator at https://paperstack.uchicago.edu/explorer.

## CODE AVAILABILITY

Data that support the findings of this study will be made available through the Qresp[70] curator at https://paperstack.uchicago.edu/explorer.

## REFERENCES

1. Gill, S. S. et al. Quantum computing: a taxonomy, systematic review and future directions. *Wiley Online Library* **52**, 66–114 (2021).
2. Hermans, S. L. N. et al. Qubit teleportation between non-neighboring nodes in a quantum network. *Nature* **605**, 663 (2022).
3. Pompili, M. et al. Realization of a multinode quantum network of remote solid-state qubits. *Science* **372**, 259–264 (2021).
4. Bradley, C. E. et al. A ten-qubit solid-state spin register with quantum memory up to one minute. *Phys. Rev. X* **9**, 031045 (2019).
5. Bourassa, A. et al. Entanglement and control of single nuclear spins in isotopically engineered silicon carbide. *Nat. Mater.* **19**, 1319–1325 (2020).
6. Nickerson, N. H., Fitzsimons, J. F. & Benjamin, S. C. Freely scalable quantum technologies using cells of 5-to-50 qubits with very lossy and noisy photonic links. *Phys. Rev. X* **4**, 041041 (2014).
7. Hensen, B. et al. Loophole-free bell inequality violation using electron spins separated by 1.3 kilometres. *Nature* **526**, 682–686 (2015).
8. Oh, D. K. et al. Top-down nanofabrication approaches toward single-digit-nanometer scale structures. *J. Mech. Sci. Technol.* **35**, 837–859 (2021).
9. Humphreys, P. C. et al. Deterministic delivery of remote entanglement on a quantum network. *Nature* **558**, 268–273 (2018).
10. Müller, M., Bounouar, S., Jöns, K. D., Glässl, M. & Michler, P. On-demand generation of indistinguishable polarization-entangled photon pairs. *Nat. Photonics* **8**, 224–228 (2014).
11. Degen, C. L., Reinhard, F. & Cappellaro, P. Quantum sensing. *Rev. Mod. Phys.* **89**, 035002 (2017).
12. Chen, X., Fu, Z., Gong, Q. & Wang, J. Quantum entanglement on photonic chips: a review. *Adv. photonics* **3**, 064002 (2021).
13. Erhard, M., Krenn, M. & Zeilinger, A. Advances in high-dimensional quantum entanglement. *Nat. Rev. Phys.* **2**, 365–381 (2020).
14. Kazimierczuk, T., Fröhlich, D., Scheel, S., Stolz, H. & Bayer, M. Giant rydberg excitions in the copper oxide $Cu_2O$. *Nature* **514**, 343–347 (2014).
15. Browaeys, A., Barredo, D. & Lahaye, T. Experimental investigations of dipole-dipole interactions between a few rydberg atoms. *J. Phys. B. Mol. Opt. Phys.* **49**, 152001 (2016).
16. Schmid, W., Nieper, U. & Weber, J. Donor-acceptor pair spectra in Si:In LPE-layers. *Solid State Commun.* **45**, 1007–1011 (1983).
17. Ziemelis, U. O. & Parsons, R. R. Sharp line donor-acceptor pair luminescence in silicon. *Can. J. Phys.* **59**, 784–801 (1981).
18. Kuwabara, H., Yamada, S. & Tsunekawa, S. Radiative recombination in β-SiC doped with boron. *J. Lumin.* **12-13**, 531–536 (1976).
19. Ivanov, I. G., Henry, A., Yan, F., Choyke, W. J. & Janzen, E. Ionization energy of the phosphorus donor in 3C-SiC from the donor-acceptor pair emission. *J. Appl. Phys.* **108**, 063532 (2010).
20. Choyke, W. J. & Patrick, L. Luminescence of donor-acceptor pairs in cubic SiC. *Phys. Rev. B* **2**, 4959 (1970).
21. Dischler, B. et al. Resolved donor-acceptor pair-recombination lines in diamond luminescence. *Phys. Rev. B* **49**, 1685–1689 (1994).
22. Freitas Jr, J. A., Klein, P. B. & Collins, A. T. Evidence of donor-acceptor pair recombination from a new emission band in semiconducting diamond. *Appl. Phys. Lett.* **64**, 2136 (1994).
23. Thomas, D. G., Gershenzon, M. & Trumbore, F. A. Pair spectra and 'edge' emission in gallium phosphide. *Phys. Rev.* **133**, 1A A269 (1964).
24. Williams, F. Donor-acceptor pairs in semiconductors. *Phys. Stat. Sol.* **25**, 493–512 (1968).
25. Dean, P. J. Bound excitons and donor-acceptor pairs in natural and synthethic diamond. *Phys. Rev.* **139**, A588–A602 (1965).
26. Dingle, R. & Ilegems, M. Donor-acceptor pair recombination in GaN. *Solid State Commun.* **9**, 175–180 (1971).
27. Tan, Q. et al. Donor-acceptor pair quantum emitters in hexagonal boron nitride. *Nano Lett.* **22**, 1331–1337 (2022).
28. Kobr, L. et al. Fast photodriven electron spin coherence transfer: A quantum gate based on a spin exchange J-jump. *J. Am. Chem. Soc.* **134**, 12430–12433 (2012).
29. Wu, Y. et al. Covalent radical pairs as spin qubits: Influence of rapid electron motion between two equivalent sites on spin coherence. *J. Am. Chem. Soc.* **140**, 13011–13021 (2018).
30. Bernien, H. et al. Probing many-body dynamics on a 51-atom quantum simulator. *Nature* **551**, 579 (2017).
31. Morgado, M. & Whitlock, S. Quantum simulation and computing with Rydberg-interacting qubits. *AVS Quantum Sci.* **3**, 23501 (2021).
32. Kamiyama, S. et al. Extremely high quantum efficiency of donor-acceptor-pair emission in N and B doped 6H-SiC. *J. Appl. Phys.* **99**, 093108 (2006).
33. Krzyaniak, M. D. et al. Fast photo-driven electron spin coherence transfer: the effect of electron-nuclear hyperfine coupling on coherence dephasing. *J. Mater. Chem. C* **3**, 7962 (2015).
34. Zhang, S. B. & Northrup, J. E. Chemical potential dependence of defect formation energies in GaAs: Application to Ga self-diffusion. *Phys. Rev. Lett.* **67**, 2339–2342 (1991).
35. Freysoldt, C., Neugebauer, J. & Van de Walle, C. G. Fully ab initio finite-size corrections for charged-defect supercell calculations. *Phys. Rev. Lett.* **102**, 1–4 (2009).
36. Heyd, J., Scuseria, G. & Ernzerhof, M. Hybrid functionals based on a screened coulomb potential. *J. Chem. Phys.* **118**, 8207–8215 (2003).
37. Perdew, J., Burke, K. & Ernzerhof, M. Generalized gradient approximation made simple. *Phys. Rev. Lett.* **77**, 3865–3868 (1996).
38. Spaldin, N. A. A beginners guide to the modern theory of polarization. *J. Solid State Chem.* **195**, 2–10 (2012).
39. Resta, R. Macroscopic polarization in crystalline dielectrics: the geometric phase approach. *Rev. Mod. Phys.* **66**, 899–915 (1994).
40. King-Smith, R. & Vanderbilt, D. Theory of polarization of crystalline solids. *Phys. Rev.* **47**, 1651–1654 (1993).
41. Pizzi, G., Vitale, V. & Arita, R. Wannier90 as a community code: new features and applications. *J. Phys. Condens. Matter* **32**, 165902 (2020).
42. Marzari, N., Mostofi, A. A., Yates, J. R., Souza, I. & Vanderbilt, D. Maximally localized wannier functions: Theory and applications. *Rev. Mod. Phys.* **84**, 1419–1475 (2012).
43. Maze, J. R. et al. Properties of nitrogen-vacancy centers in diamond: the group theoretic approach. *N. J. Phys.* **13**, 025025 (2011).

44. Anderson, C. P. et al. Electrical and optical control of single spins integrated in scalable semiconductor devices. *Science* **366**, 1225–1230 (2019).
45. Dederichs, P., Blügel, S., Zeller, R. & Akai, H. Ground states of constrained systems: Application to cerium impurities. *Phys. Rev. Lett.* **53**, 2512–2515 (1984).
46. Kaduk, B., Kowalczyk, T. & Van Voorhis, T. Constrained density functional theory. *Chem. Rev.* **112**, 321–370 (2012).
47. Williams, F. E. Theory of the energy levels of donor-acceptor pairs. *J. Phys. Chem. Lett.* **12**, 265–275 (1960).
48. Ruhoff, P. T. Recursion relations for multi-dimensional franck-condon overlap integrals. *Chem. Phys.* **186**, 355–374 (1994).
49. Alkauskas, A., Lyons, J. L., Steiauf, D. & Van de Walle, C. G. First-principles calculations of luminescence spectrum line shapes for defects in semiconductors: The example of GaN and ZnO. *Phys. Rev. Lett.* **109**, 267401 (2012).
50. Alkauskas, A., McCluskey, M. D. & Van de Walle, C. G. Tutorial: Defects in semiconductors - combining experiment and theory. *J. Appl. Phys.* **119**, 181101 (2016).
51. Jin, Y. et al. Photoluminescence spectra of point defects in semiconductors: Validation of first-principles calculations. *Phys. Rev. Mater.* **5**, 084603 (2021).
52. Gali, A. Recent advances in the ab initio theory of solid-state defect qubits. *Nanophotonics* **12**, 359–397 (2023).
53. Stoneham, A. M. *Theory of Defects in Solids: Electronic Structure of Defects in Insulators and Semimetals* (Oxford University Press, 1975).
54. Huang, K. & Rhys, A. Theory of light absorption and non-radiative transitions in F-centers. *Proc. R. Soc.* **204**, 406–423 (1950).
55. Schneider, H., Dischler, B., Wild, C. & Koidl, P. Intrinsic radiative lifetimes of donor-acceptor pair excitations in diamond. *Phys. Rev. B* **51**, 677–680 (1995).
56. Gali, A. Ab initio theory of the nitrogen-vacancy center in diamond. *Nanophotonics* **8**, 1907–1943 (2019).
57. Alkauskas, A., Dreyer, C. E., Lyons, J. L. & Van de Walle, C. G. Role of excited states in shockley-read-hall recombination in wide-band-gap semiconductors. *Phys. Rev. B* **93**, 1–5 (2016).
58. Davidsson, J. Theoretical polarization of zero phonon lines in point defects. *J. Phys. Condens. Matter* **32**, 385502 (2020).
59. Anderson, C. P. et al. Five-second coherence of a single spin with single-shot readout in silicon carbide. *Sci. Adv.* **8**, eabm5912 (2022).
60. Thomas, D. G., Jopfield, J. J. & Augustyniak, W. M. Kinetics of radiative recombination at randomly distributed donors and acceptors. *Phys. Rev.* **140**, 202–220 (1965).
61. Giannozzi, P., Baseggio, O. & Bonfa, P. Quantum ESPRESSO toward the exascale. *J. Chem. Phys.* **152**, 154105 (2020).
62. Giannozzi, P., Andreussi, O. & Brumme, T. Advanced capabilities for materials modelling with quantum ESPRESSO. *J. Phys. Condens. Matter* **29**, 465901 (2017).
63. Giannozzi, P., Baroni, S. & Bonini, N. Quantum ESPRESSO: A modular and open-source software project for quantum simulations of materials. *J. Phys. Condens. Matter* **21**, 395502 (2009).
64. Govoni, M. & Galli, G. Large scale GW calculations. *J. Chem. Theory Comput.* **11**, 2680 (2015).
65. Yu, V. W. & Govoni, M. GPU acceleration of large-scale full-frequency GW calculations. *J. Chem. Theory Comput.* **18**, 4690 (2022).
66. Schlipf, M. & Gygi, F. Optimization algorithm for the generation of ONCV pseudopotentials. *Comput. Phys. Commun.* **196**, 36–44 (2015).
67. Hamann, D. R. Optimized norm-conserving vanderbilt pseudopotentials. *Comput. Phys. Commun.* **88**, 1–10 (2013).
68. Shikata, S. et al. Precise measurements of diamond lattice constant using bond method. *Jpn. J. Appl. Phys.* **57**, 11 (2018).
69. Taylor, A. & Jones, R. M. *Silicon Carbide - A High Temperature Semiconductor*, (eds O'Connor, J. R. & Smiltens, J.) 147 (Pergamon Press, Oxford, London, New York, 1960).
70. Govoni, M. et al. Qresp, a tool for curating, discovering and exploring reproducible scientific papers. *Sci. Data* **6**, 190002 (2019).
71. Volpe, P., Pernot, J. & Muret, P. High hole mobility in boron doped diamond for power device applications. *Appl. Phys. Lett.* **94**, 10–13 (2009).
72. Czelej, K., Śpiewak, P. & Kurzydłowski, K. Electronic structure of substitutionally doped diamond: Spin-polarized, hybrid density functional theory analysis. *Diam. Relat. Mater.* **75**, 146–151 (2017).
73. Ferrari, A. M., Salustro, S., Gentile, F. S. & Mackrodt, W. C. Substitutional nitrogen atom in diamond. a quantum mechanical investigation of the electronic and spectroscopic properties. *Carbon* **134**, 354–365 (2018).
74. Deák, P., Aradi, B., Kaviani, M., Frauenheim, T. & Gali, A. Formation of nv centers in diamond: A theoretical study based on calculated transitions and migration of nitrogen and vacancy related defects. *Phys. Rev. B* **89**, 075203 (2014).
75. Weber, J. R. et al. Quantum computing with defects. *Proc. Natl Acad. Sci. USA* **107**, 8513–8518 (2010).
76. Heremans, F. J., Fuchs, G. D. & Wang, C. F. Generation and transport of photo-excited electrons in single-crystal diamond. *Appl. Phys. Lett.* **94**, 10–13 (2009).
77. Farrer, R. G. On the substitutional nitrogen donor in diamond. *Solid State Commun.* **7**, 685–688 (1969).
78. Goss, J. P., Briddon, P. R., Jones, R. & Sque, S. Donor and acceptor states in diamond. *Diam. Relat. Mater.* **13**, 684 (2004).
79. Kalish, R. The search for donors in diamond. *Diam. Relat. Mater.* **10**, 1749 (2001).
80. Rao, M. V., Griffiths, P. & Holland, O. W. Al and B ion-implantations in 6H and 3C-SiC. *J. Appl. Phys.* **77**, 2479 (2004).
81. Freitas Jr, J. A., Bishop, S. G., Nordquist Jr, P. E. R. & Gipe, M. L. Donor binding energies determined from temperature dependence of photoluminescence spectra in undoped and aluminum-doped beta SiC films. *J. Appl. Phys.* **52**, 1695–1697 (1988).
82. Freitas Jr, J. A., Klein, P. B. & Bishop, S. G. Optical studies of donors and acceptors in cubic SiC. *Mater. Sci. Eng.* **B11**, 21–25 (1992).
83. Alkauskas, A., Buckley, B. B., Awschalom, D. D. & Van de Walle, C. G. First-principles theory of the luminescence lineshape for the triplet transition in diamond NV centres. *N. J. Phys.* **16**, 073026 (2014).

## ACKNOWLEDGEMENTS
We thank Professors Ivan G. Ivanov, Tien Son Nguyen and Wolfgang J. Choyke for many fruitful discussions. Additionally, we are grateful to Dr. Bernhard Dischler, Dr. Christoph Wild, Dr. Eckhard Wörner, Dr. Peter Koidl, and Diamond Materials for their insights. We acknowledge Robert Shreiner for his assistance with sample fabrication. This work was supported by AFOSR Grant No. FA9550-22-1-0370. We acknowledge support from Boeing through Chicago Quantum Exchange, and the computational materials science center Midwest Integrated Center for Computational Materials (MICCoM). This research used resources of the National Energy Research Scientific Computing Center (NERSC), a DOE Office of Science User Facility supported by the Office of Science of the U.S. Department of Energy under Contract No. DE-AC02-05CH11231. We acknowledge funding from NSF QLCI for Hybrid Quantum Architectures and Networks (NSF award 2016136). This work also made use of the shared facilities at the University of Chicago Materials Research Science and Engineering Center, supported by the National Science Foundation under award number DMR-2011854, as well as the resources provided by the University of Chicago's Research Computing Center.

## AUTHOR CONTRIBUTIONS
A.B. performed all DFT calculations. I.H. and J.E. designed the experimental setup and obtained PL spectra from diamond samples. Y.J. provided useful discussions for calculating PL spectra. H.B., A.H. and G.G designed and supervised the research. A.B. and G.G wrote the manuscript.

## COMPETING INTERESTS
The authors declare no competing interests.

## ADDITIONAL INFORMATION
**Supplementary information** The online version contains supplementary material available at https://doi.org/10.1038/s41524-023-01190-6.

**Correspondence** and requests for materials should be addressed to Anil Bilgin or Giulia Galli.

**Reprints and permission information** is available at http://www.nature.com/reprints

**Publisher's note** Springer Nature remains neutral with regard to jurisdictional claims in published maps and institutional affiliations.